\documentclass[slac_one]{revtex4}
\usepackage{graphicx}
\usepackage{fancyhdr}
\RequirePackage{xspace}

\usepackage{relsize}
\def\babar{\mbox{\slshape B\kern-0.1em{\smaller A}\kern-0.1em
    B\kern-0.1em{\smaller A\kern-0.2em R}}}

\def\epem       {\ensuremath{e^+e^-}\xspace}

\def\mumu       {\ensuremath{\mu^+\mu^-}\xspace}

\def\ellell     {\ensuremath{\ell^+ \ell^-}\xspace}

\def\piz   {\ensuremath{\pi^0}\xspace}

\def\pip   {\ensuremath{\pi^+}\xspace}
\def\pim   {\ensuremath{\pi^-}\xspace}

\def\Kbar  {\kern 0.2em\overline{\kern -0.2em K}{}\xspace}

\def\Kz    {\ensuremath{K^0}\xspace}
\def\Kzb   {\ensuremath{\Kbar^0}\xspace}
\def\KzKzb {\ensuremath{\Kz \kern -0.16em \Kzb}\xspace}
\def\Kp    {\ensuremath{K^+}\xspace}
\def\Km    {\ensuremath{K^-}\xspace}

\def\KpKm  {\ensuremath{\Kp \kern -0.16em \Km}\xspace}
\def\KS    {\ensuremath{K^0_{\scriptscriptstyle S}}\xspace}

\def\Kstar   {\ensuremath{K^*}\xspace}

\def\Dbar    {\kern 0.2em\overline{\kern -0.2em D}{}\xspace}

\def\Dz      {\ensuremath{D^0}\xspace}
\def\Dzb     {\ensuremath{\Dbar^0}\xspace}
\def\DzDzb   {\ensuremath{\Dz {\kern -0.16em \Dzb}}\xspace}
\def\Dp      {\ensuremath{D^+}\xspace}
\def\Dm      {\ensuremath{D^-}\xspace}

\def\DpDm    {\ensuremath{\Dp {\kern -0.16em \Dm}}\xspace}

\def\B       {\ensuremath{B}\xspace}
\def\Bbar    {\kern 0.18em\overline{\kern -0.18em B}{}\xspace}

\def\BB      {\ensuremath{B\Bbar}\xspace}
\def\Bz      {\ensuremath{B^0}\xspace}
\def\Bzb     {\ensuremath{\Bbar^0}\xspace}
\def\BzBzb   {\ensuremath{\Bz {\kern -0.16em \Bzb}}\xspace}
\def\Bu      {\ensuremath{B^+}\xspace}
\def\Bub     {\ensuremath{B^-}\xspace}

\def\BpBm    {\ensuremath{\Bu {\kern -0.16em \Bub}}\xspace}

\def\BorBbar    {\kern 0.18em\optbar{\kern -0.18em B}{}\xspace}
\def\DorDbar    {\kern 0.18em\optbar{\kern -0.18em D}{}\xspace}
\def\KorKbar    {\kern 0.18em\optbar{\kern -0.18em K}{}\xspace}

\def\jpsi     {\ensuremath{{J\mskip -3mu/\mskip -2mu\psi\mskip 2mu}}\xspace}
\def\psitwos  {\ensuremath{\psi{(2S)}}\xspace}

\mathchardef\Upsilon="7107
\def\Y#1S{\ensuremath{\Upsilon{(#1S)}}\xspace}

\def\FourS {\Y4S}

\mathchardef\Deltares="7101
\mathchardef\Xi="7104
\mathchardef\Lambda="7103
\mathchardef\Sigma="7106
\mathchardef\Omega="710A

\def\Deltabar{\kern 0.25em\overline{\kern -0.25em \Deltares}{}\xspace}
\def\Lbar{\kern 0.2em\overline{\kern -0.2em\Lambda\kern 0.05em}\kern-0.05em{}\xspace}
\def\Sigbar{\kern 0.2em\overline{\kern -0.2em \Sigma}{}\xspace}
\def\Xibar{\kern 0.2em\overline{\kern -0.2em \Xi}{}\xspace}
\def\Obar{\kern 0.2em\overline{\kern -0.2em \Omega}{}\xspace}
\def\Nbar{\kern 0.2em\overline{\kern -0.2em N}{}\xspace}
\def\Xb{\kern 0.2em\overline{\kern -0.2em X}{}\xspace}

\def\mes        {\mbox{$m_{\rm ES}$}\xspace}

\def\DeltaE     {\mbox{$\Delta E$}\xspace}

\newcommand{\tev}{\ensuremath{\mathrm{\,Te\kern -0.1em V}}\xspace}
\newcommand{\gev}{\ensuremath{\mathrm{\,Ge\kern -0.1em V}}\xspace}
\newcommand{\mev}{\ensuremath{\mathrm{\,Me\kern -0.1em V}}\xspace}
\newcommand{\kev}{\ensuremath{\mathrm{\,ke\kern -0.1em V}}\xspace}
\newcommand{\ev}{\ensuremath{\mathrm{\,e\kern -0.1em V}}\xspace}
\newcommand{\gevc}{\ensuremath{{\mathrm{\,Ge\kern -0.1em V\!/}c}}\xspace}
\newcommand{\mevc}{\ensuremath{{\mathrm{\,Me\kern -0.1em V\!/}c}}\xspace}
\newcommand{\gevcc}{\ensuremath{{\mathrm{\,Ge\kern -0.1em V\!/}c^2}}\xspace}
\newcommand{\mevcc}{\ensuremath{{\mathrm{\,Me\kern -0.1em V\!/}c^2}}\xspace}

\def\mus  {\ensuremath{\rm \,\mus}\xspace}

\def\mus        {\ensuremath{\,\mu{\rm s}}\xspace}    

\def\to                 {\ensuremath{\rightarrow}\xspace}

\def\pep2{PEP-II}

\def\gsim{{~\raise.15em\hbox{$>$}\kern-.85em
          \lower.35em\hbox{$\sim$}~}\xspace}
\def\lsim{{~\raise.15em\hbox{$<$}\kern-.85em
          \lower.35em\hbox{$\sim$}~}\xspace}

\def\CP                {\ensuremath{C\!P}\xspace}

\xspace

\newcommand{\epjBase}        {Eur.\ Phys.\ Jour.\xspace}

\newcommand{\jprBase}        {Phys.\ Rev.\xspace}
\newcommand{\jplBase}        {Phys.\ Lett.\xspace}
\newcommand{\nimBaseA}       {Nucl.\ Instrum.\ Methods Phys.\ Res., Sect.\ A\xspace}

\newcommand{\nimBaseC}       {Nucl.\ Instrum.\ Methods Phys.\ Res., Sect.\ C\xspace}

\newcommand{\npBase}         {Nucl.\ Phys.\xspace}
\newcommand{\zpBase}         {Z.\ Phys.\xspace}

\newcommand{\epjc}      [1]  {\epjBase\ C~{\bf #1}}

\newcommand{\mpl}       [1]  {{Mod.\ Phys.\ Lett.\ {\bf #1}}}

\newcommand{\nim}       [1]  {\nimBaseC~{\bf #1}}

\newcommand{\nima}      [1]  {\nimBaseA~{\bf #1}}

\newcommand{\npb}       [1]  {\npBase\ B~{\bf #1}}

\newcommand{\npbps}     [1]  {{Nucl.\ Phys.\ B~Proc.\ Suppl.\ {\bf #1}}}

\newcommand{\plb}       [1]  {\jplBase\ B~{\bf #1}}

\newcommand{\pr}        [1]  {\jprBase\ {\bf #1}}

\newcommand{\progtp}    [1]  {{Prog.\ Theor.\ Phys.\ {\bf #1}}}

\newcommand{\zpc}       [1]  {\zpBase\ C~{\bf #1}}

\def\jetset74   {\mbox{\tt Jetset \hspace{-0.5em}7.\hspace{-0.2em}4}\xspace}

\newcommand{\gevcccc}{\ensuremath{{\mathrm{\,Ge\kern -0.1em V^2\!/}c^4}}\xspace}

\def\K {\ensuremath{K}\xspace}

\def\Kmaybestar {\ensuremath{K^{(*)}\xspace}}
\def\afb {\mbox{${\cal A}_{FB}$}\xspace}

\def\ctk {\ensuremath{\cos\theta_{K}}\xspace}

\def\ctl {\ensuremath{\cos\theta_{\ell}}\xspace}
\def\ctlsq {\ensuremath{\cos^2\theta_{\ell}}\xspace}
\def\emu       {\ensuremath{e^+\mu^-}\xspace}
\def\fl {\mbox{$F_L$}\xspace}

\def\kll {\B\to\Kmaybestar\ellell\xspace}
\def\kllshort {(K,K^{*})\ellell\xspace}

\def\modekavgee {\ensuremath{B\to K\epem}\xspace}

\def\modekavgll {\ensuremath{B\to K\ellell}\xspace}
\def\modekavgmm {\ensuremath{B\to K\mumu}\xspace}

\def\modekll {\ensuremath{B^+\rightarrow K^+\ellell}\xspace}

\def\modekstee {\ensuremath{B\rightarrow K^{*}\epem}\xspace}

\def\modekstll {\ensuremath{B\rightarrow K^{*}\ellell}\xspace}
\def\modekstmm {\ensuremath{B\rightarrow K^{*}\mumu}\xspace}

\def\mue       {\ensuremath{\mu^+e^-}\xspace}

\def\modekavgllshort {\ensuremath{K\ellell}\xspace}

\def\modekstllshort {\ensuremath{K^{*}\ellell}\xspace}

\def\cseven {C_{7}}
\def\cnine {C_{9}}
\def\cten {C_{10}}

\begin{document}

\title{
        {\mathversion{bold}
         Angular and Rate Asymmetries in the Decays $\kll$}
}

%

\author{Kevin T. Flood, on behalf of the \babar\, Collaboration}
\affiliation{University of Wisconsin, Madison, WI 53706, USA}

\begin{abstract}
We use a sample of 384 million $\BB$ decays collected with the \babar\,
detector at the \pep2\ asymmetric $\epem$ storage ring to study the
flavor-changing neutral current decays $\kll$, where $\ellell$
is either $\epem$ or $\mumu$. We present measurements in two
dilepton mass bins, one below the $\jpsi$ resonance and the other above,
of the lepton forward-backward asymmetry $\afb$ and the longitudinal
$K^{*}$ polarization $F_L$ in $\modekstll$, along with
isospin rate asymmetries in $\modekstll$ and $\modekavgll$ final states.
\end{abstract}

\maketitle

\thispagestyle{fancy}

The decays $\kll$, where $\ellell$ is either $\epem$ or $\mumu$,
arise from flavor-changing neutral current processes
that are forbidden at tree level in the Standard Model (SM).
The lowest-order SM processes contributing to these decays
are the photon penguin, $Z$ penguin and $W^+W^-$
box diagrams.
Their amplitudes are expressed
in terms of hadronic form factors and effective Wilson coefficients
$\cseven$, $\cnine$ and $\cten$,
representing the electromagnetic penguin diagram, and
the vector part and the axial-vector part of the $Z$ penguin and
$W^+W^-$ box diagrams, respectively~\cite{Buchalla}.
New physics contributions may enter the penguin and box diagrams
at the same order as the SM diagrams~\cite{Ali:2002jg}.
We present measurements of the lepton forward-backward
asymmetry $\afb$ and longitudinal $K^{*}$ polarization $F_L$ in
$\modekstll$, along with
isospin rate asymmetries in $\modekstll$ and $\modekavgll$ final states,
in two bins of dilepton mass squared $q^2=m_{\ellell}^2$, one below the
$\jpsi$ resonance and the other above.
Sensitive indirect searches for new physics effects using these
observables~\cite{NewPhysics} are possible as hadronic
uncertainties in calculations are expected to cancel~\cite{Kruger:1999xa}.
\babar\, results on branching fractions, direct $\CP$ and lepton-flavor
asymmetries, $\afb$ and $F_L$ have been previously published, however,
only a limit for $\afb$ in the low $q^2$ region was established~\cite{oldbabar}.

The \CP-averaged isospin asymmetry
\begin{eqnarray}
A^{K^{(*)}}_{I} \equiv
\frac
{{\cal B}(\Bz \to K^{(*)0}\ellell) - r {\cal B}(\B^{\pm} \to K^{(*)\pm}\ellell)}
{{\cal B}(\Bz \to K^{(*)0}\ellell) + r {\cal B}(\B^{\pm} \to K^{(*)\pm}\ellell)}
\end{eqnarray}
\noindent
where $r = \tau_0/\tau_+=1/(1.07\pm 0.01)$ is the
ratio of the $B^0$ and $B^+$ lifetimes~\cite{hfag},
has a SM expectation of $+6-13\%$ as $q^2 \to 0 $ for $\modekstll$~\cite{Feldmann:2002iw}.
This is consistent with the measured asymmetry of 3$\pm$3\% in $B \to K^*\gamma$~\cite{hfag}.
A calculation of the predicted $K^{*+}$ and $K^{*0}$ SM rates integrated over the low $q^2$
region gives $A^{K^{*}}_{I} = -0.00^{+0.005}_{-0.006}$~\cite{beneke05,Feldmann:priv2008}. Given that
the expected SM isospin asymmetry arises from pure photon penguin contributions to $\modekstll$,
there is no expectation of such an asymmetry in $\modekavgllshort$.
In the high $q^2$ region, although there may be possible contributions
from higher charmonium resonances, the measured asymmetry in $\jpsi\Kmaybestar$
is only a few percent~\cite{hfag}, and any SM $A^{K^{*}}_{I}$
can be reasonably expected to be similarly insignificant.
We measure isospin asymmetries in a low $q^2$ region $0.1 < q^2 < 7.02\gevcccc$, and a high
region $q^2>10.24\gevcccc$, where any likely contributions from
$J/\psi$ and $\psi(2S)$ have been removed by vetoing events $7.02<q^2<10.24\gevcccc$ and
$12.96<q^2<14.06\gevcccc$, respectively.

The $K^*$ longitudinal polarization fraction $\fl$ can be
determined from the distribution of the angle
$\ctk$ between the $K$ and the $B$ directions in the $K^*$ rest frame
using a fit to $\ctk$ of the form~\cite{KrugerMatias}
\begin{equation}
\frac{3}{2} \fl \cos^2\theta_K + \frac{3}{4}(1-\fl)(1-\cos^2\theta_K)
\end{equation}
Likewise, the lepton forward-backward asymmetry $\afb$ can be
determined from the distribution of the angle $\ctl$ between
the $\ell^+(\ell^-)$ and the $B(\Bbar)$ direction in
the $\ellell$ rest frame using a fit to $\ctl$ of the form~\cite{KrugerMatias}
\begin{equation}
{{3}\over{4}}\fl (1-\ctlsq) + {{3}\over{8}}(1-\fl )(1+\ctlsq) + \afb \ctl
\end{equation}
These angular measurements are also done in low and high $q^2$ regions,
with the low region slightly more narrowly defined than above, $0.1 < q^2 < 6.25\gevcccc$,
in order to absolutely remove any possible contributions from $J/\psi$ and
$\psi(2S)$ decays, which have a distinct angular structure but are
otherwise indistinguishable from signal decays.
As shown in Eq.~3, an experimental determination of $\fl$ is
required to obtain a model-independent $\afb$ result.

\begin{figure}[!h]
\includegraphics[width=0.48\linewidth]{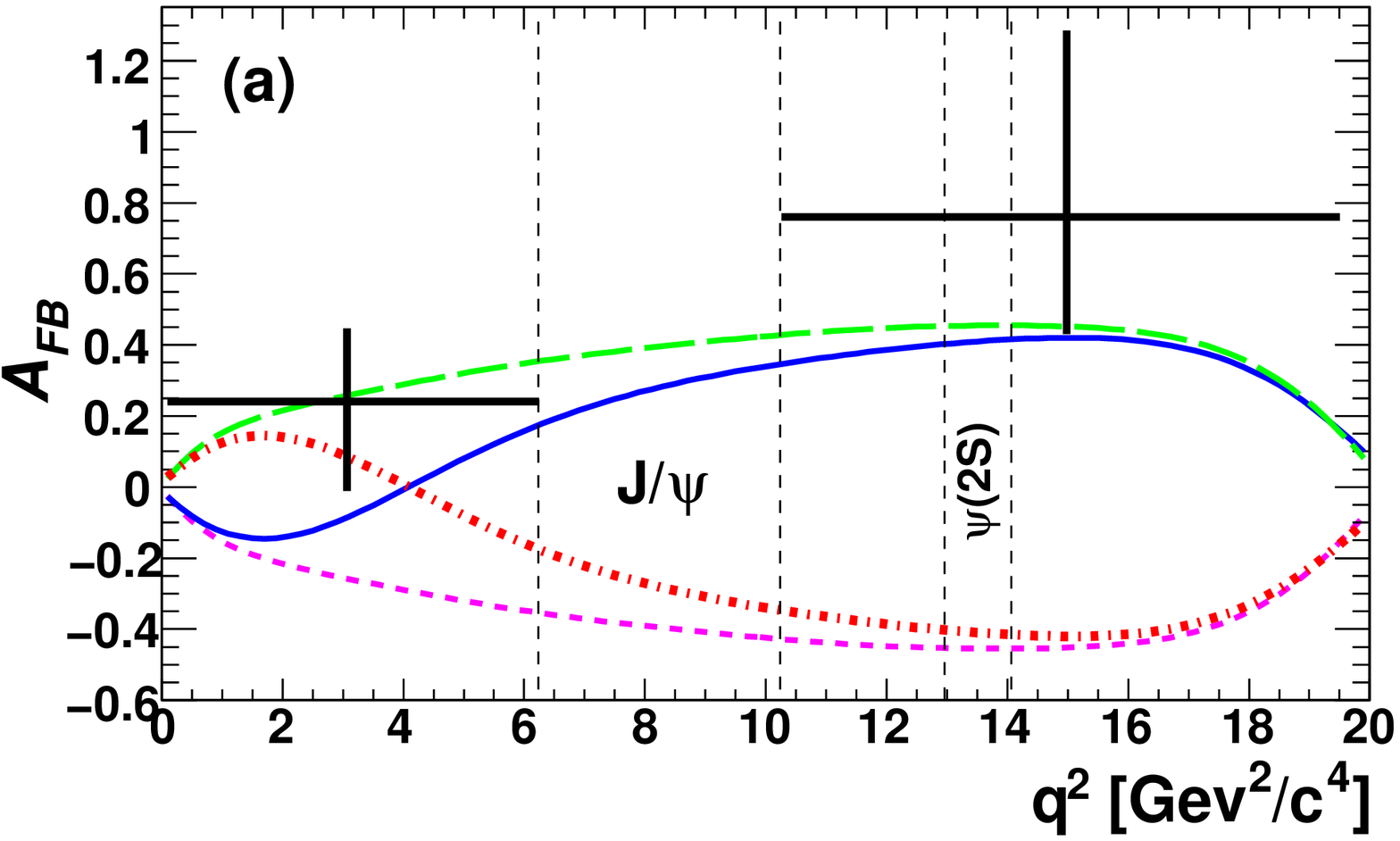}\includegraphics[width=0.48\linewidth]{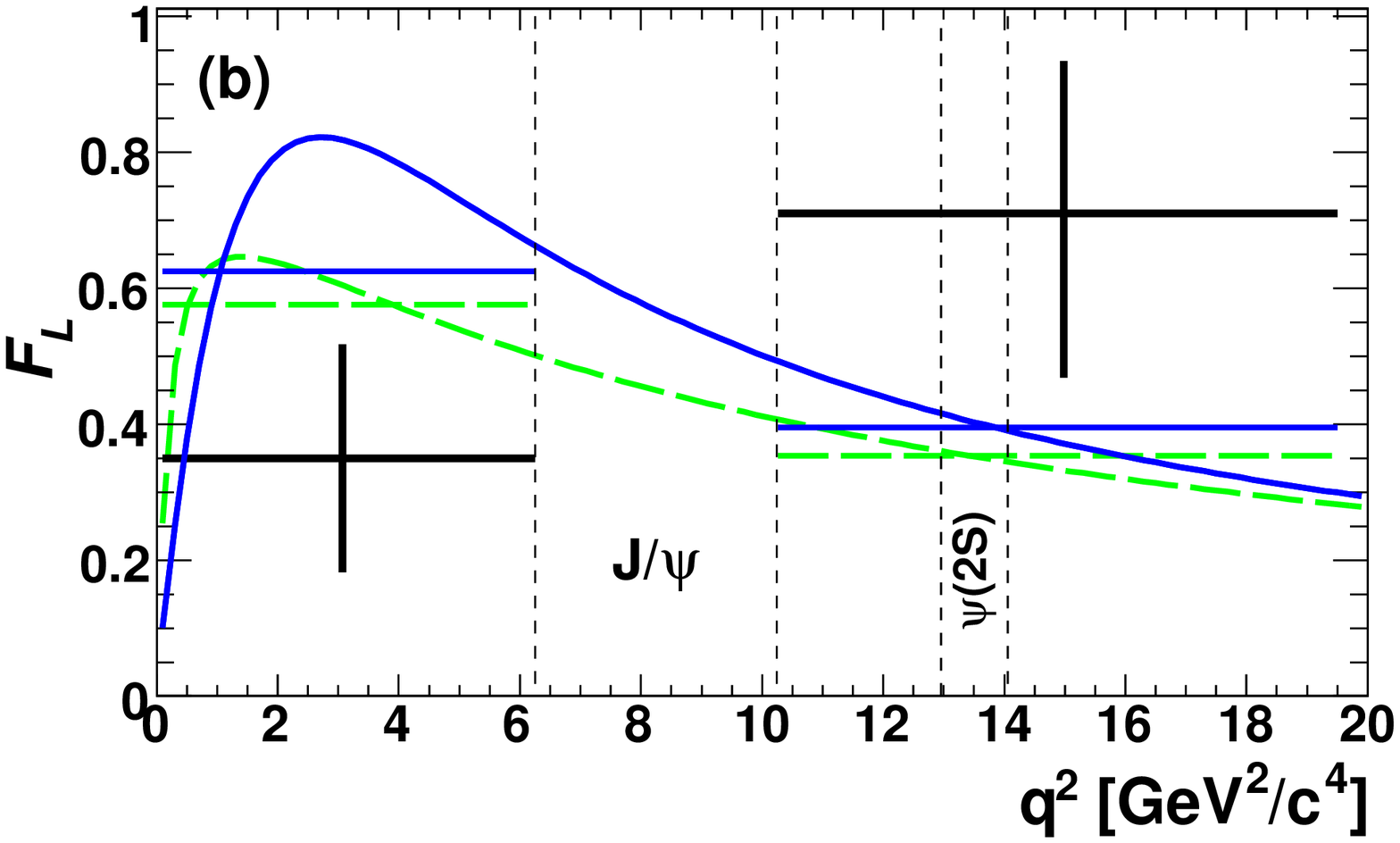}
\caption{(a) $\afb$ and (b) $F_{L}$ results showing comparisons
with SM (solid), $\cseven = -\cseven^{SM}$ (long dash),
$\cnine \cten = -\cnine^{SM} \cten^{SM}$ (short dash),
$\cseven = -\cseven^{SM}, \cnine \cten = -\cnine^{SM} \cten^{SM}$ (dash-dot).
Statistical and systematic errors are added in quadrature.
Expected $F_{L}$ values integrated over each $q^2$ region are also shown.
The $F_{L}$ curves with $\cnine \cten = -\cnine^{SM} \cten^{SM}$
are nearly identical to the two curves shown.}
\label{fig:newafbfl}
\end{figure}

Variations in $\fl$ and $\afb$ as a function of $q^2$ result from
interference among the different amplitudes. The expected
SM behavior of $\fl$ and $\afb$, along with variations
due to non-SM opposite-sign Wilson coefficients, is shown by
the curves in Fig.~\ref{fig:newafbfl}.
At low $q^2$, where $\cseven$ dominates, $\afb$ is expected to be small
with a zero-crossing point at $q^2 \sim 4 \gevcccc$~\cite{Ali01,AFB_SM,bsgamma_beneke}.
An experimental constraint on the magnitude of $\cseven$ comes
from the measured inclusive branching fraction for
$b\to s\gamma$~\cite{hfag,bsgamma_beneke,bsgamma}, which
corresponds to the limit $q^2 \to 0$, but its sign is unknown.
At high $q^2$ in the SM, the product $\cnine\cten$
is expected to give a large positive asymmetry.
Right-handed weak currents would have an opposite-sign $\cnine\cten$,
leading to a negative $\afb$ contribution at high $q^2$.
Contributions from other non-SM processes could change the magnitudes and relative signs
of $\cseven$, $\cnine$ and $\cten$, and introduce complex phases between
them~\cite{KrugerMatias,Hovhannisyan}.

We use a data sample of $384$ million $\BB$ pairs
collected at the $\FourS$ resonance
with the \babar\ detector~\cite{BaBarDetector} at
the \pep2\ asymmetric-energy $\epem$ collider at SLAC.
Tracking is provided
by a five-layer silicon vertex tracker (SVT) and a 40-layer drift chamber (DCH)
in a 1.5-T magnetic field. We identify electrons
with a CsI(Tl) electromagnetic calorimeter, and muons
using an instrumented magnetic flux return.
Electrons (muons) are required to have momenta $p > 0.3 (0.7) \gevc$
in the laboratory frame.
We combine photons with electrons when they are consistent with bremsstrahlung,
and do not use electrons that are associated with a photon
converting to a low-mass $\epem$ pair.
We identify $\Kp$ using a detector of internally reflected
Cherenkov light, as well as ionization energy loss
measurements from the DCH and SVT.
Charged tracks other than identified $e$, $\mu$ and $K$ candidates are
treated as pions. Neutral pion candidates are formed from
two photons with laboratory energies $E_{\gamma} > 50 \mev$ and an invariant mass
between $115$ and $155 \mevcc$.

We reconstruct signal events in ten separate final states
containing an $\epem$ or $\mumu$ pair, and a $\KS(\to \pip\pim)$, $\Kp$,
or $\Kstar(892)$ candidate with an invariant mass $0.82 < M(K\pi) < 0.97 \gevcc$.
We reconstruct $K^{*0}$ candidates in the
final state $K^+\pi^-$, and $K^{*+}$ candidates in the final states
$K^+\pi^0$ and $\KS\pi^+$ (charge conjugation is implied throughout).
Neutral $\KS \to \pip \pim$ candidates are required to have an invariant mass consistent
with the nominal $K^0$ mass~\cite{PDG}, and a flight distance from the primary interaction point
which is more than three times its uncertainty.
We also study final states $K^{(*)}h^{\pm}\mu^{\mp}$,
where $h$ is a track with no particle identification (PID) requirement applied,
to characterize backgrounds from hadrons misidentified as muons.
Signal decays are characterized using the kinematic
variables $\mes=\sqrt{s/4 -p^{*2}_B}$ and
$\Delta E = E_B^* - \sqrt{s}/2$, where $p^*_B$ and $E_B^*$ are
the $B$ momentum and energy in the $\Upsilon(4S)$ center-of-mass (CM) frame,
and $\sqrt{s}$ is the total CM energy.
We define a fit region $\mes > 5.2 \gevcc$, with
$-0.07<\Delta E<0.04$ ($-0.04<\Delta E<0.04$) $\gev$ for
$e^+e^-$ ($\mu^+\mu^-$) final states in the low $q^2$ region, and
$-0.08<\Delta E<0.05$ ($-0.05<\Delta E<0.05$) $\gev$ for high $q^2$.

The main backgrounds arise from random combinations of
leptons from semileptonic $B$ and $D$ decays, which are suppressed
using event shape variables, vertexing information and missing energy
combined in event selection neural networks (NNs).
We use simulated samples of signal and background events in
the construction of the NNs and, assuming rates consistent with
accepted values~\cite{hfag}, we optimize the NN selections
for best statistical significance in the number of expected
signal events in our dataset.
A further background contribution comes from
$B \to D(\rightarrow \Kmaybestar \pi) \pi$ decays
where both pions are misidentified as leptons.
This background is significant only in dimuon final states,
as the pion misidentification rate for electrons is $<0.1\%$.
We veto these events by assigning the pion mass to a muon candidate and
requiring that the invariant mass of the hypothetical $\Kmaybestar\pi$ system
be outside the range 1.84-1.90$\gevcc$.
We perform blind fits for both the isospin and angular observables, in which all
event selection criteria were determined prior to examining events in the fit region.


We directly fit the data with $A_{I}^{\Kmaybestar}$ as a floating parameter
using a simultaneous unbinned maximum likelihood $\mes$ fit across all modes
contributing to a particular $A_{I}$ measurement. An ARGUS shape~\cite{ArgusShape}
with floating shape parameter and normalization is used to describe combinatorial background.
For the signal, we use a fixed Gaussian shape unique to each final state,
with mean and width determined from fits to the analogous
final states in the large samples of vetoed $J/\psi\Kmaybestar$ events.
Events with misidentified muons escaping the $D$ mass veto are accounted for using
$\Kmaybestar h^{\pm}\mu^{\mp}$ events weighted by the per-particle
probability, determined from uncorrelated PID control samples,
for $h^{\pm}$ to be misidentified.
We also account for small contributions from misreconstructed signal
events and charmonium events that escape the charmonium mass vetos.
We test the fit using the large samples of vetoed
$J/\psi \Kmaybestar$ and $\psitwos \Kmaybestar$ events, and
find good agreement with accepted values for branching fractions to the
individual final states used here and the small charmonium isospin asymmetries~\cite{PDG}.

We consider systematic uncertainties associated with reconstruction efficiencies;
hadronic background parameterization in di-muon final states;
peaking background contributions obtained from simulated events;
and possible isospin asymmetries in the background pdfs.
We quantify efficiency-related systematics using the vetoed
$\jpsi\Kmaybestar$ samples, including charged track,
$\piz$, and $\KS$ reconstruction, PID, NN selection,
and the $\DeltaE$ and $\K^{*}$ mass selections.
Nearly all systematic effects largely cancel in the
$A_I$ ratio, and the final systematic uncertainties
are small compared to the statistical ones.

\begin{figure}
\begin{center}
\includegraphics[width=0.55\textwidth]{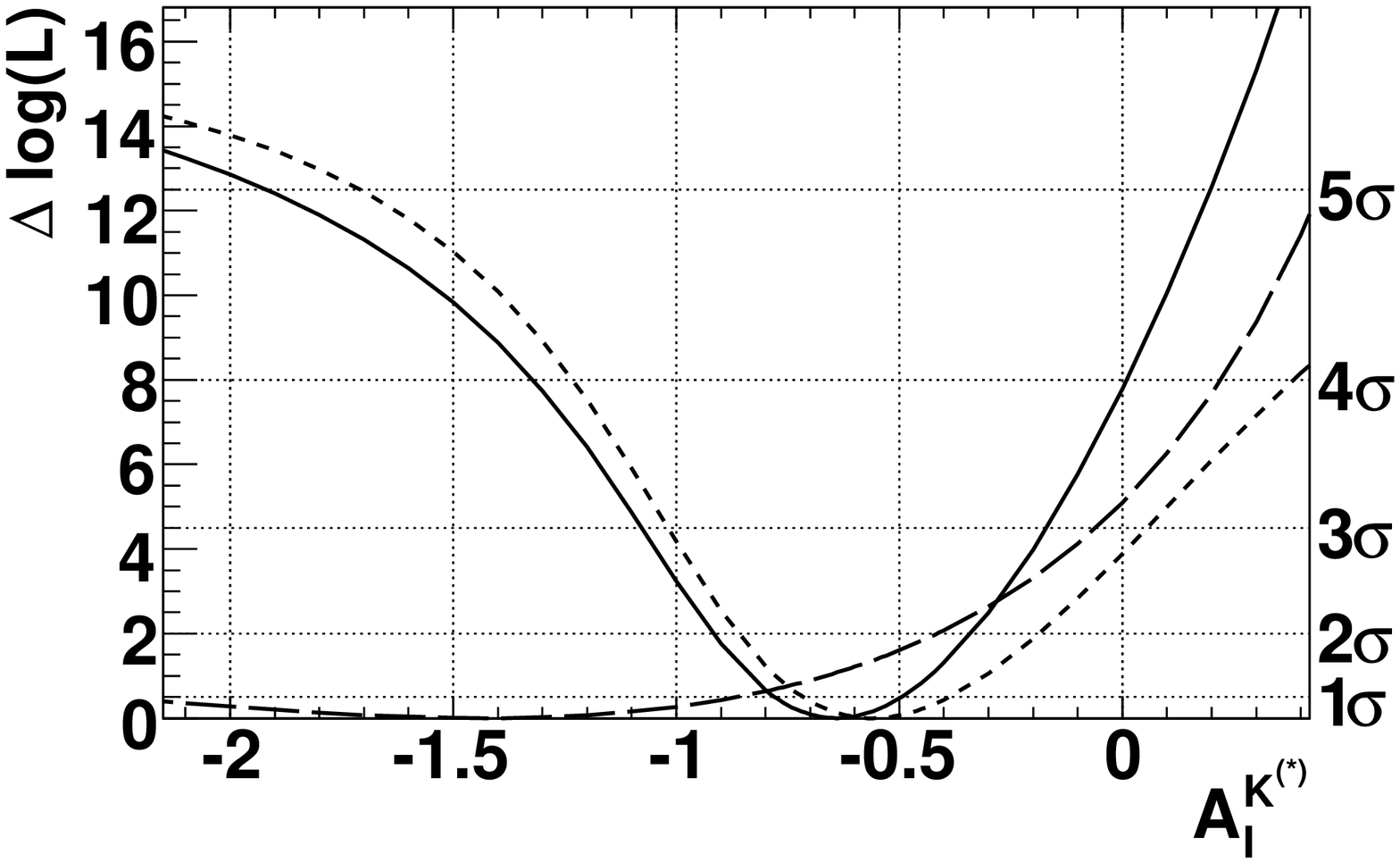}
\caption{Low $q^2$ region $A_{I}^{\Kmaybestar}$ fit likelihood curves.
$\modekavgllshort$ [long dash],
$\modekstllshort$ [short dash],
$\kllshort$ [solid].}
\label{fig:nllscans}
\end{center}
\end{figure}

Table~\ref{tab:isoressys} shows the $A_{I}^{\Kmaybestar}$ results.
We find no significant isospin asymmetries in the high $q^2$ region, but
we find evidence for large negative asymmetries in the low region.
We calculate the statistical significance with which a null
isospin asymmetry hypothesis is rejected using the change in
log likelihood $\sqrt{2 \Delta \ln{\cal L}}$ between the nominal
fit to the data and a fit with $A_{I}^{\Kmaybestar}=0$ fixed.
Figure~\ref{fig:nllscans} shows the likelihood curves obtained
from the $\modekavgllshort$ and $\modekstllshort$
fits. The parabolic nature of the curves in the $A_{I}^{\Kmaybestar}>-1$ region
demonstrates the essentially Gaussian nature of our fit results
in the physical region, and the right-side axis of
Figure~\ref{fig:nllscans} shows purely statistical significances
based on Gaussian coverage.
Incorporating the relatively small systematic uncertainties as
a scaling factor on the change in log likelihood, the significance
in the low $q^2$ region that $A_{I}^{\Kmaybestar}$
is different from zero is $3.2\sigma$ for $\modekavgllshort$
and $2.7\sigma$ for $\modekstllshort$.
We have verified these confidence intervals by performing fits to ensembles
of simulated datasets generated with $A_{I}^{\Kmaybestar}=0$ fixed, and we find
frequentist coverage consistent with the $\Delta \ln{\cal L}$ calculations.
The highly negative $A_{I}^{\Kmaybestar}$ values for both $\modekavgllshort$ and $\modekstllshort$
at low $q^2$ suggest that this asymmetry may be insensitive to the hadronic final state,
and so we sum the likelihood curves as shown in Figure~\ref{fig:nllscans} and obtain
$A_{I}^{\Kmaybestar} = -0.64^{+0.15}_{-0.14} \pm 0.03$. Including systematics, this is a $3.9 \sigma$
difference from a null $A_{I}^{\Kmaybestar}$ hypothesis.

\begin{table}
\centering
\caption{$A_{I}^{\Kmaybestar}$ results. Errors are statistical and
systematic, respectively.}
{\footnotesize
\begin{tabular}{lccc}
\hline \hline
Mode          & low $q^2$                                  & high $q^2$                    \\ \hline
$\modekavgmm$ & $-0.91_{\mathrm{-\infty}}^{+1.2}\pm0.18$   & $0.39_{-0.46}^{+0.35}\pm0.04$ \\
$\modekavgee$ & $-1.41_{-0.69}^{+0.49} \pm 0.04$           & $0.21_{-0.41}^{+0.32}\pm0.03$ \\
$\modekavgll$ & $-1.43_{-0.85}^{+0.56} \pm 0.05$           & $0.28_{-0.30}^{+0.24}\pm0.03$ \\
$\modekstmm$  & $-0.26_{-0.34}^{+0.50} \pm0.05$            &$-0.08_{-0.27}^{+0.37}\pm0.05$ \\
$\modekstee$  & $-0.66_{-0.17}^{+0.19} \pm0.02$            & $0.32_{-0.45}^{+0.75}\pm0.03$ \\
$\modekstll$  & $-0.56_{-0.15}^{+0.17} \pm0.03$            & $0.18_{-0.28}^{+0.36}\pm0.04$ \\
\hline \hline
\end{tabular}
}
\label{tab:isoressys}
\end{table}

For the angular analysis, because of the relatively small number of expected
$\modekstllshort$ signal candidates in each $q^2$ region, a simultaneous
fit over $\mes$, $\ctk$ and $\ctl$ is not possible, and an iterated
fitting procedure is used.
The first step is an $\mes$ fit, with the same underlying components as above,
in each $q^2$ region for total signal ($N_S$) and combinatoric background ($N_B$)
yields for the combination of all $\modekstllshort$ final states.
The second fit is to $\ctk$ for events with $\mes > 5.27 \gevcc$, where
the only free parameter is $\fl$, and normalizations for signal and combinatorial
background events are taken from the initial $\mes$ fit.
The background normalization is obtained by integrating,
for $\mes > 5.27 \gevcc$, the ARGUS shape resulting from the $\mes$ fit.
We model the $\ctk$ shape of the combinatorial background
using $\epem$ and $\mumu$ events, as well as lepton-flavor violating $\emu$
and $\mue$ events, in the $5.20 < \mes < 5.27 \gevcc$ sideband.
Simulated events are used to account for the small remaining background contributions.
The signal distribution given in Eq.~2 is folded with a model for the signal
acceptance as a function of $\ctk$ obtained from simulated signal events.

The final fit is to $\ctl$, again for events with $\mes > 5.27 \gevcc$,
where the only free parameter is $\afb$. This fit requires the value
of $\fl$ as an input, which is taken from the second fit, as are the
normalizations for signal and combinatorial background.
The $\ctl$ shape of the combinatorial background is obtained
using the same sideband samples as for the $\ctk$ fit.
Correlated leptons coming from $B\to D^{(*)}\ell\nu$,
$D\to K^{(*)}\ell\nu$ give rise to a peak in the combinatorial
background at $\ctl>0.7$ which varies as a function of $\mes$, and
we consider this variation in our study of systematic uncertainties.
As for $\ctl$, the signal distribution given in Eq.~3 is folded with a
model of signal acceptance as a function of $\ctl$ taken from simulated events.
We again test our fits using the large sample of vetoed charmonium events,
where $\afb$ is expected to be zero, and branching fractions and the
$K^*$ polarization are well-known~\cite{JPsi_BF, JPsi_FL}, and obtain
results consistent with accepted values for each of the six individual
$\modekstllshort$ final states, as well as their combination.
We further test our methodology by performing the
$\mes$ and $\ctl$ fits on a $\modekll$ sample,
where $\afb \sim 0$ is expected in most new physics models as well as the SM,
and find values consistent with a null result in both low and high $q^2$ regions.

We consider systematic effects from several sources.
Uncertainties in yields due to variations in the
ARGUS shape in the $\mes$ fits are propagated into both angular fits, and
uncertainties on the fit $\fl$ values are propagated into the $\afb$ fits.
Combinatorial background angular shapes are varied by dividing the sideband sample
into two disjoint regions in $\mes$.
We vary the signal model using simulated events generated with
different form factors~\cite{AFB_SM, BallZwicky} and a wide range
of values of $\cseven$, $\cnine$ and $\cten$. We perform fits
to large ensembles of datasets for each generator variation
to determine an average absolute fit bias.
Finally, we constrain the modeling of mis-reconstructed signal events
from the fits to the charmonium samples, where it is the largest
systematic uncertainty. As with $A_{I}^{\Kmaybestar}$, the final
systematic uncertainties are small compared to the statistical ones.

The results for $\fl$ and $\afb$ are shown in Fig.~\ref{fig:newafbfl}.
In the low $q^2$ region, where we expect $\afb=-0.03 \pm 0.01$~\cite{Huber:2007vv}
and $\fl=0.63 \pm 0.03$~\cite{KrugerMatias} from the SM,
we measure $\afb=0.24^{+0.18}_{-0.23} \pm 0.05$
and $\fl=0.35 \pm 0.16 \pm 0.04$, where the first error
is statistical and the second is systematic.
In the high $q^2$ region, the SM expectation is
$\afb \sim 0.26$ and $\fl \sim 0.40$,
and we measure $\afb=0.76^{+0.52}_{-0.32} \pm 0.07$
and $\fl=0.71^{+0.20}_{-0.22} \pm 0.04$,
with a signal yield of $36.6 \pm 9.6$ events.
Theoretical uncertainties on the expected SM $\fl$ and $\afb$ values
in the high $q^2$ region are difficult to characterize,
and the quoted values are obtained from our
signal event generator~\cite{AFB_SM, BallZwicky}.

The magnitude of possible contributions from new physics to $C_{10}$ can
be constrained by a positive sign of $\afb$ at high $q^2$.
By combining $\afb$ with inclusive branching fraction results,
an upper bound of $|C_{10}^{NP}| < \sim 7$ can be obtained,
improving on an upper bound derived solely from
branching fraction results of $|C_{10}^{NP}| < \sim 10$~\cite{Bobeth:2008ij}.
The $\afb$ results additionally exclude a wrong-sign $C_9C_{10}$
from purely right-handed weak currents at more
than 3 standard deviations significance.
The low $q^2$ AFB result suggests that a zero-crossing point, a
distinctive SM feature, may not be present.
Some angular asymmetries in the $\K^*$ system
include the longitudinal polarization as a component and
are more sensitive to possible new physics contributions than
measurements of $\fl$ alone, as predicted values for these
asymmetries in the SM and various new physics models can be
calculated with relatively small theory uncertainties~\cite{Lunghi:2006hc}.
However, such asymmetries could not be considered here as they
require analysis of a dataset substantially greater than
currently available.
Our results are consistent with measurements by
Belle~\cite{Belle_afb}, and replace the earlier
\babar\ results in which only a lower limit on $\afb$ was set in
the low $q^2$ region~\cite{oldbabar}.

\end{document}